\documentclass[11pt]{article}%
\usepackage[conf]{ametsoc}  
\usepackage{graphicx}
\usepackage{amsmath}
\usepackage{amsfonts}
\usepackage{amssymb}
\usepackage{fullpage}
\usepackage{tabularx}
\usepackage{subcaption}
\usepackage{booktabs}
\usepackage{appendix}
\usepackage{xurl}
\usepackage{tikz}
\usepackage{perpage}
\MakePerPage{footnote}
\providecommand{\U}[1]{\protect\rule{.1in}{.1in}}
\begin{document}
\tolerance=9999 \pretolerance=9999
\title{Custom Loss Functions in Fuel Moisture Modeling}
\author{
    \begin{tabular}[t]{c}
        Jonathon Hirschi \\
        University of Colorado Denver \\
        \\
        Advisor: Jan Mandel
    \end{tabular}
}

\date{July 22, 2024}
\maketitle

\begin{center}
    \includegraphics[width=0.3\textwidth]{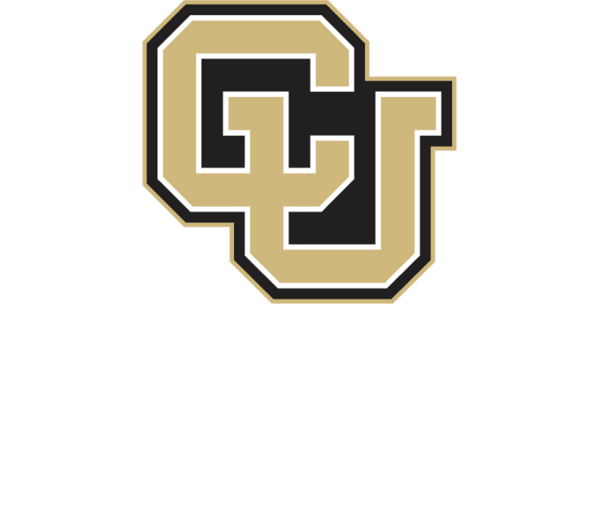}

    A Project submitted in partial fulfillment\\ of the requirements for the degree of\\ Master of Science in Statistics

    \vspace{1em}

    \textbf{Committee Members:}\\
    Dr. Jan Mandel\textsuperscript{1}, Advisor\\
    Dr. Joshua French\textsuperscript{1}\\
    Dr. Kyle Hilburn\textsuperscript{2}\\

    \textsuperscript{1}University of Colorado Denver\\
    \textsuperscript{2}Cooperative Institute for Research in the Atmosphere, Colorado State University\\

    \vspace{10em}
    This research was partially supported by NASA grants 80NSSC23K1344, 80NSSC23K1118, 80NSSC22K1405, 80NSSC22K1717, and 80NSSC19K1091.
    
\end{center}

\newpage

\begin{abstract}
Fuel moisture content (FMC) is a key predictor for wildfire rate of spread (ROS). Machine learning models of FMC are being used more in recent years, augmenting or replacing traditional physics-based approaches. Wildfire rate of spread (ROS) has a highly nonlinear relationship with FMC, where small differences in dry fuels lead to large differences in ROS. In this study, custom loss functions that place more weight on dry fuels were examined with a variety of machine learning models of FMC. The models were evaluated with a spatiotemporal cross-validation procedure to examine whether the custom loss functions led to more accurate forecasts of ROS. Results show that the custom loss functions improved accuracy for ROS forecasts by a small amount. Further research would be needed to establish whether the improvement in ROS forecasts leads to more accurate real-time wildfire simulations.
\end{abstract}

\textbf{Key Terms and Abbreviations}
\begin{itemize}
    \item \textbf{FMC:} Fuel Moisture Content, a ratio interpretable as a percentage of the weight of water in a stick relative to the weight of the stick.
    \item \textbf{ROS:} Fire Rate of Spread, units of meters per second.
    \item \textbf{ML:} machine learning.
    \item \textbf{Response Variable:} the variable that is being modeled, or the output of the model.
    \item \textbf{Predictors:} data variables used to predict the response variable. Also known as features, covariates, etc.
    \item \textbf{Residual:} the difference between the predicted values of a model and the observed response value.
\end{itemize}

\newpage
\tableofcontents
\newpage

\section{Background}

The goal of this project is to modify the way that machine learning models of FMC are typically developed in order to produce more accurate predictions of wildfire ROS. Custom loss functions that place more weight on dry fuels will be proposed. These loss functions are hypothesized to improve the accuracy of machine learning models at predicting wildfire ROS. In this section, the importance of wildfire modeling and the key inputs to the models will be discussed. Then there will be a discussion of background information on machine learning models and loss functions.

\subsection{Wildfires and Climate Change}

Wildfires impact our lives in many ways, including destruction of property and public health effects. Toxic gases and harmful particulate matter emitted from wildfires pose a threat to air quality and human health \citep[e.g.,][]{WHO-2024-WFS}. Wildfires also affect water quality by depositing metals and other toxins into waterways and contributing to erosion and flooding \citep{USGS-2024-WQA}. The U.S. Congress Joint Economic Committee estimated that total monetary costs of wildfires as of 2022 is between \$394 billion and \$893 billion per year after accounting for numerous effects from wildfires, including burned structures, smoke impacts, watershed damage, etc. \citep{JEC-2023-CEW}.

Climate change is projected to increase the frequency and severity of wildfires, caused primarily by higher temperatures and less frequent precipitation. There is clear evidence that the changes predicted by climate models are already underway. Since the early 1980s, there has been a drastic increase in the number, size, and intensity of wildfires throughout the Western United States \citep{Cartier-2022-UFQ}. The total burned area has increased, and more areas of the world are susceptible to fires \citep{IPCC-2021-LCI}. The three largest fires in Colorado’s history all occurred in 2020 \citep{CDFPC-2024-HWI}. Wildfires are also more frequent overall and more common throughout the year, even during months that traditionally were wetter and less prone to ignition. The Intergovernmental Panel on Climate Change (IPCC) reports that “fire weather season has already lengthened by 18.7\% globally between 1979 and 2013” \citep{IPCC-2021-LCI}. Traditionally, the months of May through September were considered Colorado’s fire season \citep{WFCA-2023-CFS}. However, the most destructive wildfire in Colorado’s history, the Marshall Fire, occurred outside the city of Boulder in December of 2021 \citep{CDFPC-2024-HWI}.

There are complicated interactions between the environment, human activity, and wildfires. Some important weather patterns make fires more prevalent and more severe when they do ignite. Drought is considered the “dominant driver of fire emissions” \citep{IPCC-2021-LCI}, because dry vegetation provides the needed fuel for fires. Both the lack of precipitation and lower air humidity contribute to dry fuels. Wind is another key factor in fire activity. Strong winds dry out fuels, making fires more likely to ignite, and once fires are underway wind feeds the fire oxygen and physically spreads the flames, thus increasing wildfire spread \citep{Richardson-2022-GIW}. Human activity can increase the prevalence and intensity of fires. Wildfires are often started by downed power lines, fireworks, or other human causes. Additionally, certain land management practices have increased the risk of large fires. Fire ``exclusion”, sometimes referred to as fire ``suppression”, refers to a land management practice where all fire is deliberately stopped and prevented. This results in a buildup of fuels which has been linked to the increasing risk of ``mega-fires” \citep{Williams-2013-EOH}. Other human activity can decrease the intensity of wildfires, such as prescribed burns which prevent the buildup of fuels that cause mega-fires.

Given the substantial impacts of wildfires, researchers are trying to understand wildfires and make predictions about how they will behave. Mathematical and statistical modeling can assist with wildfire prevention by identifying areas that are susceptible to major fires. This information can be used to provide wildfire risk warnings to the public. Also, this can help focus resources used for controlled burns or other mitigation efforts in the areas where it is needed most. Finally, when wildfires are actively burning, models can inform where the fire will spread and where harmful emissions might travel. Accurate models of wildfire dynamics can therefore help save money and save lives.

\subsection{Fuel Moisture Content}

Fuel moisture content (FMC) is a critical component of wildfire susceptibility and fire spread. It is a measure of the water content in vegetation, and is expressed as a percentage of the ``dry weight" of the fuel \citep{NCEI-2024-DFM}. The dry weight is what the fuel would weigh when entirely dried out of moisture. In practice, sticks of wood are dried in kilns and weighed to determine the dry weight. An FMC of 0\% would be completely dry fuel, and FMC values can range over 100\% if the total weight of the water is greater than the dry weight of the plant material. The formula for FMC is:

\begin{equation}
    \text{FMC} (\%) = \frac{\text{Fuel Weight} - \text{Dry Weight}}{\text{Dry Weight}} \cdot 100\%
\end{equation}

Dry fuels burn more readily, and wetter fuels burn more slowly or not at all. Atmospheric conditions affect FMC in a variety of ways. Temperature is a key component of FMC; warmer temperatures dry out fuels, with other factors being held constant. Relative humidity (RH) is a measure of water content in the air, and is another key component of FMC; more moisture in the air generally leads to more moisture in fuels. Higher wind speeds lead to drier fuels. Finally, solar radiation also dries out fuels. Cloudy days with little direct solar radiation will generally lead to higher FMC conditions \citep{NWCG-2024-FWP}. It is important to note that these atmospheric conditions relate to each other in complicated ways. For example, higher temperatures can cause lower RH and are also associated with stronger solar radiation, all of which lead to lower FMC. However, higher temperatures can cause more evaporation and lead to increases in precipitation, which in turn increases FMC.

Researchers typically model dead FMC, or the water content of dead vegetation, separately from the FMC of living plants. This is due to several factors. Living plants try to maintain homeostasis, so they resist large fluctuations in moisture caused by atmospheric conditions. The water content of dead fuel, on the other hand, responds readily to atmospheric conditions and there can be large fluctuations in FMC that are relevant to wildfire spread. Additionally, live fuels are thought of as composites of various fuel types. A sagebrush, for example, has woody material of various diameters as well as herbaceous material. These various materials retain moisture in different ways and respond to fire differently \citep{NWCG-2024-FWP}. In practice, the dead FMC models are used as an input into the live FMC forecasts. In wildfire modeling software platforms like WRF-SFIRE and elsewhere, the dead FMC model is adjusted to provide estimates of live FMC based on biological factors of plant material.

Dead fuel moisture is divided into different classes based on how quickly the material responds to changes in atmospheric conditions. These include 1-hour, 10-hour, 100-hour, and 1,000-hour fuels \citep{NCEI-2024-DFM}. Table \ref{tab:classes} shows the sizes of these fuel classes. A 1-hour fuel, very small twigs for example, will get wetter quickly when exposed to precipitation, while a 1,000-hour fuel, such as a fallen log, will take much longer to absorb precipitation and thus the FMC changes more slowly. The fuel classes are defined by the lag time which determines how quickly the fuel responds to a change in atmospheric conditions. A 10-hour class fuel approaches a theoretical equilibrium moisture as $\exp(-t/10)$, where $t$ is the time in hours.

\begin{table}[ht]
\centering
\caption{Dead FMC Fuel Classes.}
\label{tab:classes}
\begin{tabular}{|l|l|}
\hline
\textbf{Time Lag}           & \textbf{Fuel Size (diameter)} \\  \hline
1-hour     & Less than 1/4 in.             \\ \hline
10-hour    & 1/4 - 1 in.             \\ \hline
100-Hour   & 1-3 in.\\ \hline
1000-Hour   & 3-8 in.\\ \hline

\end{tabular}
\end{table}

An important concept in FMC modeling is ``equilibrium moisture content". The equilibrium moisture content is an important predictor of FMC in a variety of physics-based and machine learning models. Theoretically, if environmental conditions are kept stable for a long period of time, the FMC of wood will equilibrate to its surroundings. The equilibrium FMC is the theoretical value that the FMC would approach over time, and it is calculated from relative humidity and temperature \citep[e.g.,][]{Mitchell-2018-CEM}. There are slightly different physical processes when fuels dry out versus when they absorb moisture, so researchers have developed the ``wetting equilibrium" and ``drying equilibrium" to establish what the equilibrium moisture content is for fuels under these two different scenarios \citep{Mandel-2014-RAA}. These variables are constructed from relative humidity RH (\%) \footnote{Note that RH is a ratio and can thus be considered unitless. In this project, RH will be considered to have units of percent for clarity.} and temperature T (K) in the following formulas:

\begin{equation}
\label{eq:equil}
\begin{split}
\text{Wetting Equilibrium} = &0.924\cdot\text{RH}^{0.679} + 0.000499\cdot\exp(0.1\cdot\text{RH}) + \\ &0.18\cdot(21.1 + 273.15 - \text{T})\cdot(1 - \exp(-0.115\cdot\text{RH})) \quad (\%)\\
\text{Drying Equilibrium} = &0.618\cdot\text{RH}^{0.753} + 0.000454\cdot\exp(0.1\cdot\text{RH}) + \\ &0.18\cdot(21.1 + 273.15 - \text{T})\cdot(1 - \exp(-0.115\cdot\text{RH}))\quad (\%)
\end{split}
\end{equation}

\subsection{Wildfire Rate of Spread}

The rate of spread (ROS) of a fire is a measure of how quickly fire propagates from a point of origin \citep[e.g.,][]{NFSC-2024-MFB}. Traditionally, this was measured in ``chains" per hour, where a chain is an antiquated unit of measurement corresponding to about 66ft. In recent years, researchers use the units meters per second. Factors affecting ROS include wind speed and direction, topographic slope, fuel density and type (e.g. fine twigs versus big logs), and FMC. The ROS is meant to characterize actively burning fires, but the concept can be applied to how quickly a theoretical fire would propagate given the fuel properties at that location. 

\begin{figure}[ht]
    \centering
    \begin{subfigure}{0.45\textwidth}
        \centering
        \includegraphics[width=\linewidth]{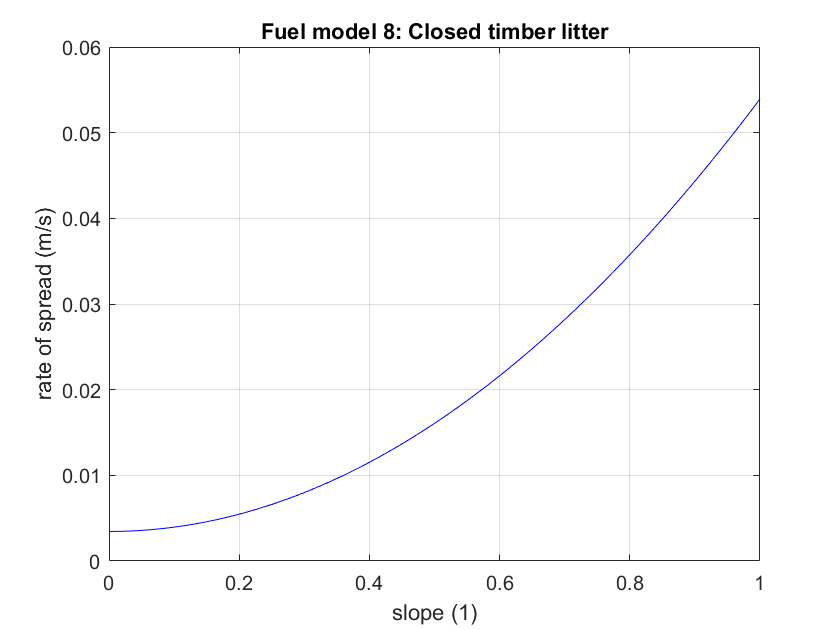}
    \end{subfigure}
    \hfill
    \begin{subfigure}{0.45\textwidth}
        \centering
        \includegraphics[width=\linewidth]{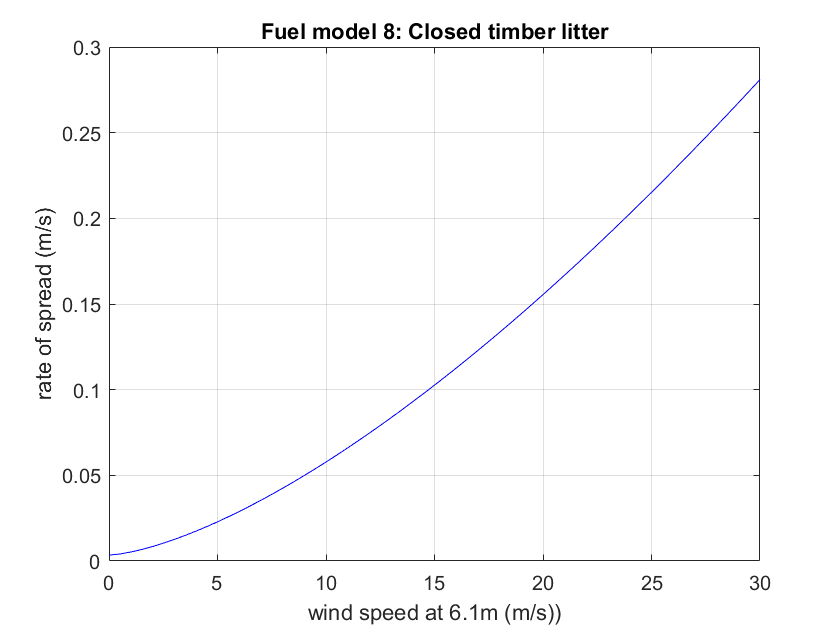}
    \end{subfigure}
    \caption{ROS vs Slope and ROS vs Wind}
    \label{fig:ros_other}
\end{figure}

Figure \ref{fig:ros_other} shows the idealized relationships between ROS and some important predictors other than FMC, while holding all else constant. These are model outputs from the physics-based tools within the wildfire modeling software project WRF-SFIRE, which will be described in greater detail later. Different fuels burn differently and will thus have different relationships between FMC and ROS. Figure \ref{fig:ros_other} shows the relationship between ROS and two key predictors: terrain slope and wind speed. For terrain slope, the units are tangent of the slope in degrees, so a value of 1 corresponds to $\text{tan} (45^\circ) = 1$  \citep{OpenWFM-2024-HTD}. The relationship is strictly increasing in both cases, and can be reasonably well-approximated as linear in certain applications.

In this project, the idealized relationship for a category of fuel known as ``closed timber litter" will be used, since it is the most representative of the widely available FMC sensors out of the available fuel categories used by researchers. Developed by Anderson in 1982, wildfire researchers often use categories of fuels, which are grouped together by environmental considerations and have characteristic burning properties \citep{NIFC-2024-FAF}. Like all fuel types used by researchers, closed timber litter is a mixture of various fuel classes. Within WRF-SFIRE, closed timber litter is composed of 71\% 10-hour dead fuels and 29\% 1-h dead fuels (0.115 $lb/ft^2$ and .046$lb/ft^2$, respectively). This is the largest ratio of pure 10-hour fuels of the 13 fuel types used within WRF-SFIRE.

\begin{figure}[p]
    \centering
    \includegraphics[width=0.8\textwidth]{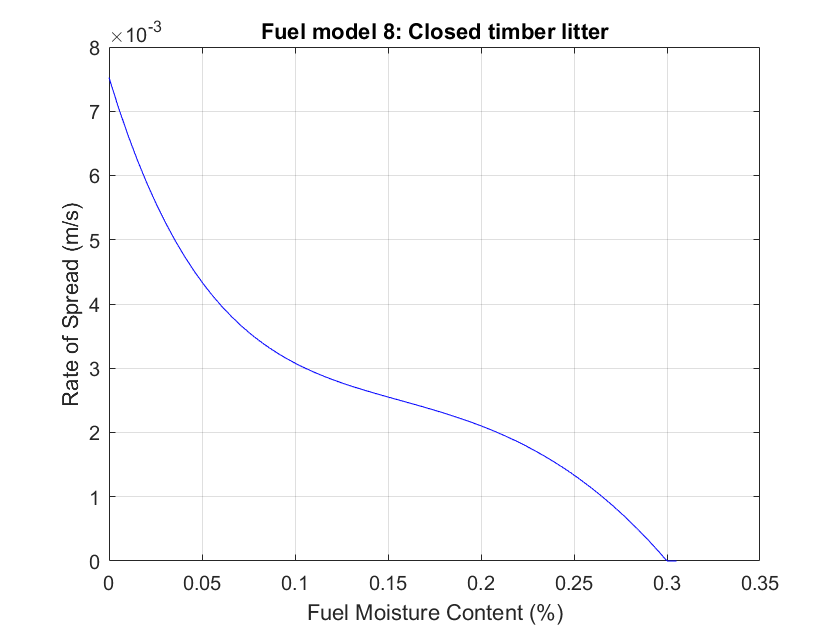}
    \caption{ROS vs FMC relationship, 0 wind.}
    \label{fig:fmc_ros_0wind}
\end{figure}

The relationship between ROS and FMC is highly nonlinear. Figure \ref{fig:fmc_ros_0wind} shows the relationship between FMC and ROS with zero wind and zero slope. Small differences in the dry region of the plot, from 0\% to around 5\%, lead to very large changes in the resulting ROS. For the wetter region of the plot, the relationship with ROS is approximately flat until it reaches a ``moisture of extinction", where fire will not spread at all. This nonlinear relationship is of primary interest in this research project. 

\begin{figure}[p]
    \centering
    \includegraphics[width=0.8\textwidth]{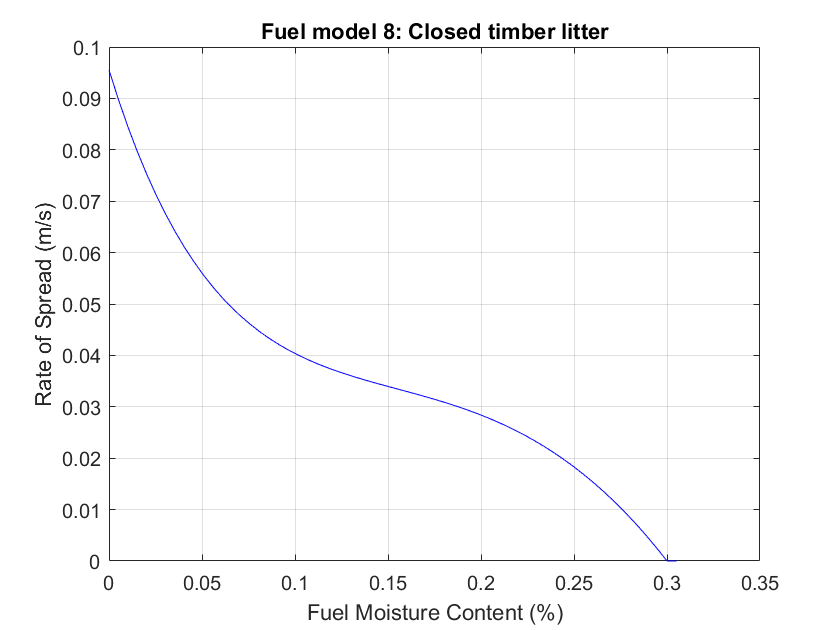}
    \caption{ROS vs FMC relationship, 3m/s wind.}
    \label{fig:fmc_ros_3wind}
\end{figure}

The relationship between FMC and ROS is much more pronounced in the presence of wind. Figure \ref{fig:fmc_ros_3wind} shows the relationship between FMC and ROS when wind speed is a constant 3m/s. Notice the values of ROS in the y-axis are much higher than in Figure \ref{fig:fmc_ros_0wind}. In the dataset collected for this project, which will be discussed in detail in Section \ref{sec:data} of this report, the average wind speed was roughly 2.9m/s. Thus, in this project the curve shown in Figure \ref{fig:fmc_ros_3wind} will be used as a function to transform FMC observations into theoretical ROS values. This is a simplification of the reality of the relationship between FMC and ROS, but the purpose of this simplification is just to develop weights for a generic custom loss function that will be applied to models of 10-hour dead FMC.

\subsection{Fuel Moisture Modeling}

\subsubsection{History and Physics-based Methods}
\hfill

Researchers studying wildfires have traditionally used physics-based models that try to represent wildfire behavior from physical first-principles, such as properties of radiative flux and heat transfer. A seminal model developed by \cite{Rothermel-1972-MMP} is a mathematical model of wildfire spread based on physical properties of heat transfer. This model includes FMC as one of the key parameters. This model has been used and built upon by wildfire researchers for decades. 

Principles of Rothermel's original model are used in WRF-SFIRE, a real-time forecasting project for wildfires which combines an extensive set of mathematical methods and data processing techniques \citep{OpenWFM-2024-HTD}. There is also a set of software tools to retrieve and format data for WRF-SFIRE called ``wrfxpy".\footnote{The code for wrfxpy is publicly available on github: \url{https://github.com/openwfm/wrfxpy}} Within wrfxpy, there is a physical model of FMC developed by \citet{Mandel-2014-RAA}, which uses theoretical physical properties of wetting and drying, combined with data assimilation through the augmented Kalman filter \citep{Vejmelka-2016-DAD}. The core differential equation in the model uses wetting equilibrium moisture, drying equilibrium moisture, and precipitation and as data inputs. The outputs of this physical model of FMC are fed into the WRF-SFIRE system for real-time wildfire simulations.

Many other physics-based models of FMC have been developed which include some combination of relative humidity, temperature, and precipitation to estimate FMC from physical principles. Also, field observations of FMC are used to tune ``parameters", or constant values, used within the physical models \citep{Catchpole-1999-EFR,Nelson-2000-PDC, vanderKamp-2017-MFS}.  Currently, the most widely used wildfire modeling tools still rely on physics-based models, and modern machine learning attempts are only beginning to impact the field within recent years.

\subsubsection{Machine Learning models of FMC}
\hfill

With modern advances in computing and more extensive observed FMC data, machine learning models of FMC have become more popular. Machine learning (ML) is a general term for statistical models that learn the patterns of observed data and are used to predict new, unobserved data. The term ``training" is used to describe the process of using data to modify model parameters. The models researchers have commonly considered within the FMC literature include various forms of linear regression, random forests, or XGBoost.\footnote{This is a commonly used abbreviation for a method called ``Extreme Gradient Boosting". See Appendix \ref{app:ml} for more detail.} Versions of these three model architectures will be used in this project, and additional details on the models can be found in the Methods section as well as Appendix \ref{app:ml}. These models are not recursive in time, meaning that the models are mapping observed weather conditions to the response variable of FMC at each instance of time without explicitly including a temporal relationship. These models will be referred to as ``static." Using derived time predictors like hour of the day and day of the year can help a static model learn time dependence, but this is different than a recursive model. Recursive models include any model where the output at time $t$ is a function of the model output at previous times. Examples of recurrent models are autoregression and Recurrent Neural Networks (RNNs), and such models will be considered in future research.

Using the previously mentioned static ML models is an active area of research in the FMC modeling literature. \cite{Lee-2020-EFM} fit various statistical models to ground station observations in South Korea, and selected a Random Forest model as the most accurate. Researchers at the National Center for Atmospheric Research (NCAR) conducted a larger study in the United States by developing and evaluated a number of ML models of FMC using weather data from the High Resolution Rapid Refresh (HRRR) weather model and NASA's VIIRS satellites \citep{McCandless-2020-EWS, Schreck-2023-MLV}. The final model in their publicly available 10-hour FMC forecasting tool is an XGBoost. 

The physical relationship between FMC and atmospheric conditions is fundamentally time-dependent. In other words, the current state of FMC depends not just on the current atmospheric conditions, but the atmospheric conditions in the past. For example, a large rain event might result in very wet fuels that take a long time to dry out. Physics-based models of FMC are time-dependent differential equations, so they are recursive. Since the underlying physical process is time dependent, it is an active area of research is to use temporally recursive models for fuel moisture modeling. \cite{Kang-2022-FMC} tested autoregressive models and a variety of neural network architecture known as Long Short-Term Memory (LSTM) to predict monthly averaged FMC observations. Other recent efforts to use recurrent neural networks (RNNs) include \cite{Fan-2021-PGD}, where an LSTM was built with the 2017 Vanderkamp physical model of FMC as an input. Additional recent efforts include \cite{Mandel-2023-BFM}, where a RNN is fit to atmospheric data from the HRRR weather model. There is not a widely available FMC model product built on recursive ML models at this time.

In this project, static ML models of FMC will be considered. Static models, which do not explicitly model the time dependence of the physical process, have weaknesses in that they struggle to learn certain patterns of a time-dependent process. However, these models have the advantage that they are much easier to build and deploy, and in practice can achieve levels of accuracy that is at least consistent with time-recursive models.

\subsection{Loss Functions in Machine Learning}

Training a ML model involves four components. 1) the dataset, 2) the ML model architecture, 3) the loss function, and 4) the algorithm used for optimizing model parameters. The parameters of a ML model are adjusted to minimize the loss function, thus the loss function controls which patterns are learned from the dataset. Loss functions are the main area of research interest in this project, and are described in this section.

\subsubsection{Loss Functions Overview}
\hfill

A loss function is intended to measure the fitting accuracy of a statistical model. Loss functions are used for training the parameters of ML models. The final model parameters that are used to make future predictions are the ones that minimize the loss function. Minimizing the loss function can be done analytically for models like linear regression, but for more complicated models it must be estimated using some variety of optimizer like gradient descent. 

When the response variable is continuous, as is the case with modeling FMC, the standard choice of loss function is the mean squared error (MSE), which is the sum of squared residuals. This is proportional to the squared L2 norm of the vector of model residuals, with the scaling factor being $1/N$. Suppose there is an observed vector of response data $y$ of length $N$ and a vector of model predictions $\hat y$. The $i^{th}$ residual of the model would be $(y_i - \hat y_i)$. The MSE for the model would thus be:

\begin{equation}
    \label{eq:mse}
    \text{MSE} = \frac{1}{N}\sum_{i=1}^N (y_i - \hat y_i)^2 = \frac{1}{N}\|y - \hat y\|^2_2
\end{equation}

There are many other potential loss functions for continuous data, such as the the mean absolute error, which is proportional to the squared L1 norm of the residuals. In this project, a number of weighted MSE loss functions will be considered. Weighted least squares minimizes the weighted residual sum of squares, with a weight $w_i$ applied to each squared residual value:

\begin{equation}
    \label{eq:wmse}
    \text{Weighted MSE} = \frac{1}{N}\sum_{i=1}^N w_i\cdot (y_i - \hat y_i)^2
\end{equation}

Weighting the loss function introduces bias into the estimator when considering the accuracy of the predictions. This addition of bias can be beneficial if the variance of the estimator is sufficiently reduced. This dynamic is referred to as the ``bias-variance tradeoff" \citep[p.\ 37]{Hastie-2010-ESL}. Weighted loss functions have been most commonly used in the context of classification problems, where the modeled outcome is a categorical variable and the loss function of choice is something like cross-entropy. When trying to build classification models on highly imbalanced datasets, such as when predicting a~very rare disease, it is often desirable to weight the loss function to give more weight to the underrepresented class. For example, in a model of a rare disease where only 1 out of 1 million observed people have the disease, a model could correctly predict 999,999 out of 1 million cases correctly by predicting that nobody ever has the disease. The model will therefore never correctly predict that somebody has the disease, so this is undesirable behavior. Weighting the loss function so that the rare cases contribute more to the total model loss can control for these errors. With continuous response variables, weighted loss functions are used for outlier detection, where iterative schemes identify influential observations and downweight them so they do not have an outsize effect on the final model fit. This is the technique used in various robust regression algorithms \citep{OLeary-1990-RRC}. In this project, weighted MSE loss functions with a continuous response variable will be considered based on purely physical considerations of the problem, not meant to address class imbalance or any outlier issues. Constructing loss functions based on physical considerations in this way is much less common in the ML literature.

\subsubsection{Custom Loss Functions in Environmental Science}
\hfill

Within wildfire modeling and broader environmental science literature, custom loss functions have been used in a number of studies, but primarily for imbalanced classification tasks as discussed previously. Researchers from the Cooperative Institute for Research in the Atmosphere (CIRA) at Colorado State University published a guide to implementing custom loss functions for environmental science applications in 2021, with a number of applications related to storm detection, generating synthetic radar imagery, and more. The paper introduces a simple exponentially weighted loss function \citep[p.\ 7]{Ebert-2021-GCL}, which will be used in this project, along with instructions to implement these loss functions in the Tensorflow software library for neural networks.

Detecting wildfires using remote sensing is an active area of research, and ML methods of image processing are popular. The problem is often framed as a classification task, where given a grid of pixels in a image, the model classifies each pixel as fire or not fire. Since the goal is early detection of wildfires, when there are fires present the number of fire pixels is very small relative to the overall number of pixels in the image. Thus, this is an imbalanced classification task and weighted loss functions are useful. \cite{Yang-2021-PFF} and \cite{Pande-2021-WSF} both use variations of convolutional neural networks (CNNs) for classification with weighted loss functions to address class imbalance. In a slightly related project, \cite{Buch-2022-SML} use a variety of neural network to predict wildfire frequency and burned area in the Western US. A custom loss function is used to get the network to minimize the loss of the model output compared to a  statistical distribution.

For custom loss functions specifically related to modeling FMC, researchers have used weighted loss functions when investigating the relationship between FMC and soil moisture. \cite{Lu-2021-EMS} use a robust regression procedure to downweight influential observations when using satellite observations to model soil moisture and live FMC. \cite{Rakhmatulina-2021-SMI} use a weighted regression scheme when trying to perform inference on the size of the effect of soil moisture on FMC. This project presents a new use of weighted loss functions in FMC modeling, with the purpose of improving ROS predictions.

\section{Data}
\label{sec:data}
\subsection{Data Acquisition}

\begin{figure}[ht]
    \centering
    \includegraphics[width=0.8\textwidth]{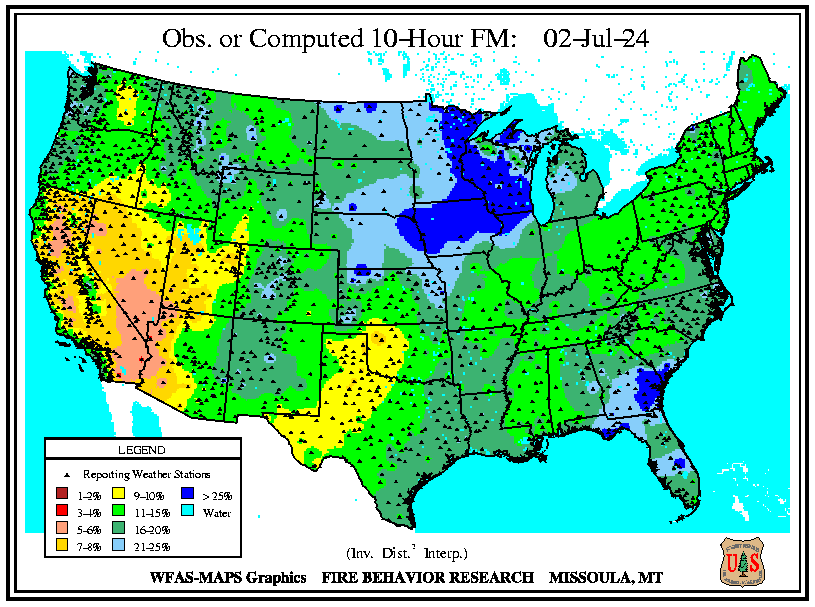}
    \caption{All RAWS FMC locations.}
    \label{fig:wfas_raws}
\end{figure}

The National Interagency Fire Center (NIFC) operates Remote Automatic Weather Stations (RAWS) across the country to measure hourly FMC and other associated environmental variables \citep{NIFC-2024-RAW}. Figure \ref{fig:wfas_raws} shows the location of all RAWS with FMC sensors. These RAWS stations\footnote{The ``S" in RAWS stands for station, so this is redundant terminology. It is an aesthetic preference that this terminology is more clear.} have sensors for 10-hour fuels, which consist of a 1/2 inch pine dowel with a moisture probe inside of it \citep{Campbell-2017-RMM}. Observations of 1-hour or 100-hour fuels are typically done through field observations. But for real-time FMC modeling, 10-hour fuels are the only class of fuel with substantial data availability. In this project, dead 10-hour FMC will be examined for this reason.

Real-time weather data from RAWS stations is acquired through Synoptic Weather API, using the Python software package ``SynoticPy" developed \citep{Synoptic-2024-SWA, Blaylock-2023-SPA}. Synoptic is an organization that grew out of the University of Utah's MesoWest program. They provide a range of atmospheric data services.

\begin{figure}[h]
    \centering
    \begin{subfigure}[b]{0.45\textwidth}
        \centering
        \includegraphics[width=\textwidth]{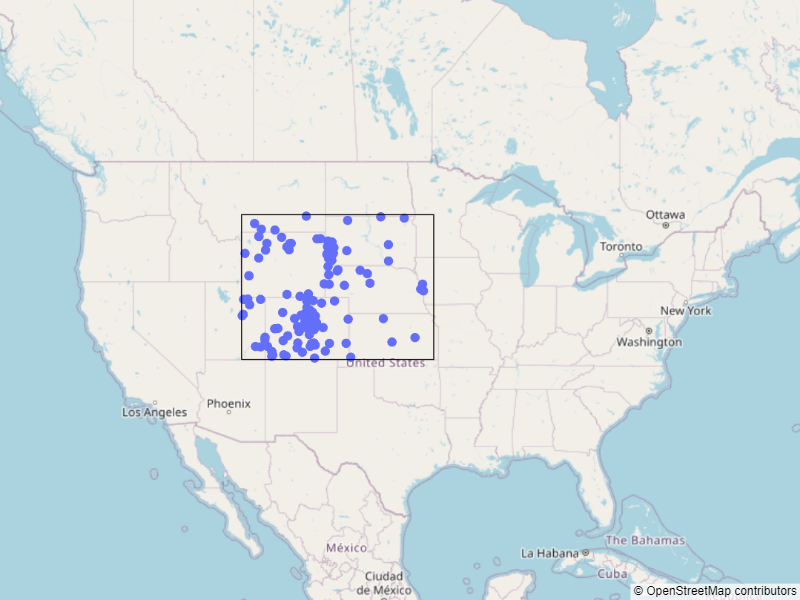}
        \caption{National View}
    \end{subfigure}
    \hspace{0.05\textwidth} 
    \begin{subfigure}[b]{0.45\textwidth}
        \centering
        \includegraphics[width=\textwidth]{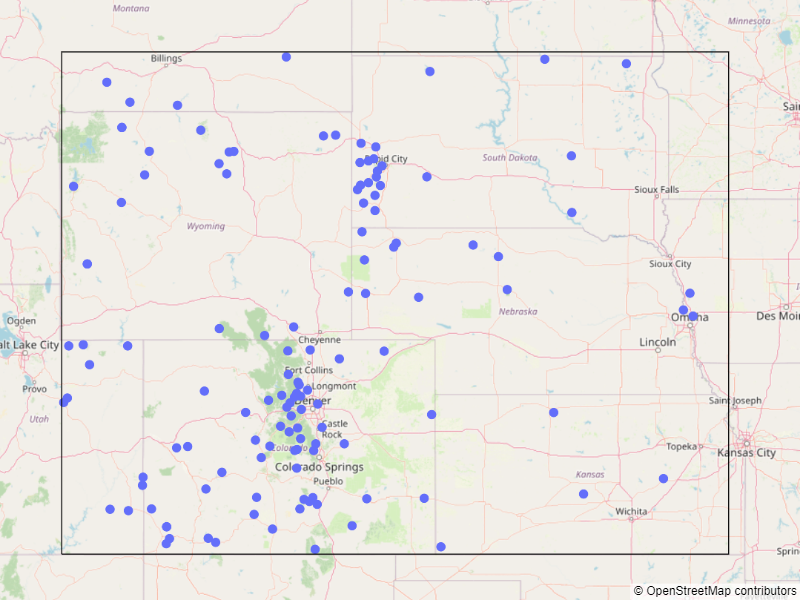}
        \caption{Close-up View}
    \end{subfigure}
    \caption{Rocky Mountain GACC - 128 RAWS Stations with Valid Data.}
    \label{fig:raws_map}
\end{figure}

For this project, 13 months of FMC and weather data was collected from RAWS stations from May 2023 through May 2024. The cross-validation procedure used in this project splits training and testing data based on time, so for the training data to cover a full year it is necessary to collect slightly more than one year of data. See Section \ref{sec:cv} for more details. The geographic area for this study was all RAWS within a spatial bounding box\footnote{
The bounding box is defined as: [Minimum Latitude, Minimum Longitude, Maximum Latitude, Maximum Longitude], and the Rocky Mountain GACC has bounding box: [$37^\circ, -111^\circ, 46^\circ, -95^\circ$]} containing the Rocky Mountain Geographic Area Coordination Center (GACC), as maintained by \cite{NIFC-2024-GAC}. The data was formatted,  filtered for erroneous data, and transformed to produce other useful predictors. It is challenging to filter out all erroneous data since RAWS fuel moisture sensors can malfunction or produce errors in ways that are not easily identified algorithmically. See Appendix \ref{app:data} for more details. Derived predictors from the data include the wetting and drying moisture equilibria, the day of year, and hour of the day from 0 to 23 UTC time. It is a simplifying assumption that hour of the day is consistent across longitude from the East to West. See Appendix \ref{app:data} for more details. In total, over one year of data was collected at 128 unique locations. After applying data filters, 875,130 total observations of 9 predictor variables were used. Figure \ref{fig:raws_map} shows the location of the 128 RAWS stations within the Rocky Mountain GACC spatial bounding box with valid data over the time period considered.

\subsection{Data Exploration}

\begin{figure}[ht]
    \centering
    \includegraphics[width=0.8\textwidth]{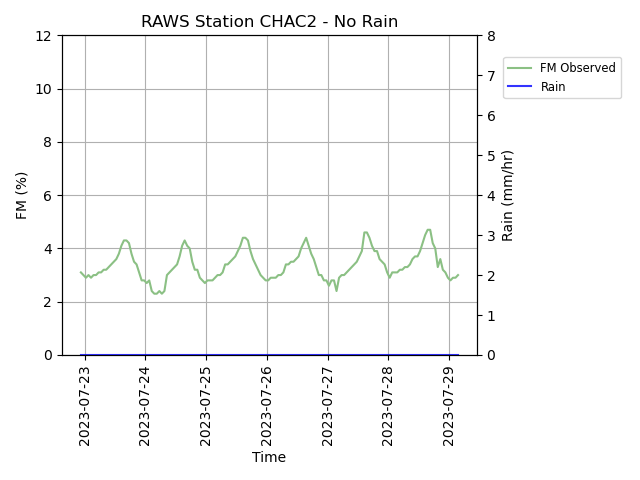}
    \caption{FMC over time, no rain.}
    \label{fig:fmc_no_rain}
\end{figure}

In the absence of precipitation, FMC exhibits a diurnal cycle. FMC is the highest at night just after when humidity is highest and temperatures are lowest. The time lag between the peak FMC and the corresponding extreme values of humidity and temperature is determined by the fuel class. So a 1-hour fuel will have peak FMC relatively shortly after the extreme humidity and temperature times, while there would be a longer time delay with a 100-hour fuel. Moisture is typically the lowest just after the heat of the day in the early afternoon when humidity is low and temperatures are high. Figure \ref{fig:fmc_no_rain} shows this pattern for a RAWS station in Colorado.\footnote{RAWS station ID CHAC2 is in southwest Colorado, at coordinates (37.19944,	-108.48917)} Using the hour of the day as a predictor can help ML models capture this cyclical pattern. 

\begin{figure}[ht]
    \centering
    \includegraphics[width=0.8\textwidth]{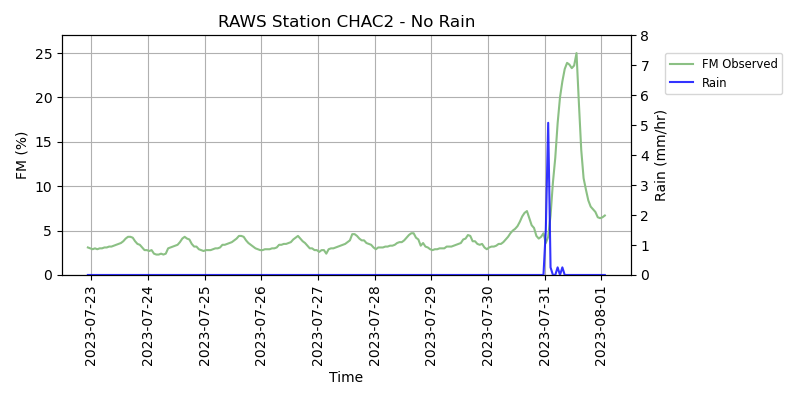}
    \caption{FMC over time, with rain.}
    \label{fig:fmc_with_rain}
\end{figure}

Precipitation causes fuels to directly absorb of water. In Figure \ref{fig:fmc_with_rain}, the FMC increases very quickly just following a rain event. Note the y-axis difference between Figures \ref{fig:fmc_no_rain} and \ref{fig:fmc_with_rain}. This pattern of sudden and rapid increases in a response variable are difficult to model.

\begin{figure}[ht]
    \centering
    \includegraphics[width=0.8\textwidth]{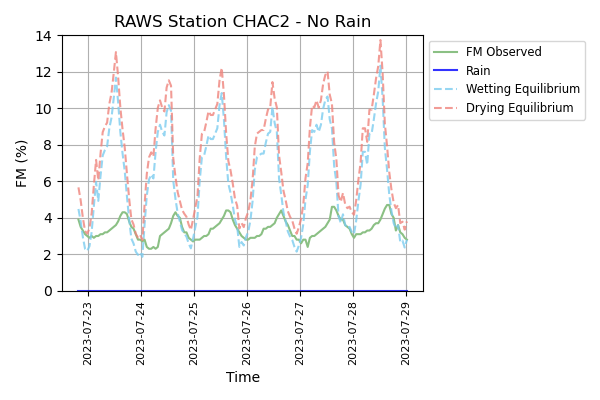}
    \caption{FMC and equilibrium over time.}
    \label{fig:eq_plot}
\end{figure}

Relative humidity and air temperature have a complicated relationship with FMC. Additionally, RH and temperature are related to each other. Temperature strongly affects how much water can be held in the air. As temperature increases, RH decreases holding all else constant. Using equilibrium moisture content as a predictor can leverage the physical understanding of the relationship between RH, temperature, and FMC. As discussed previously, the equilibrium moisture content is the theoretical FMC of a piece of wood if kept in constant atmospheric conditions for a long time. Equilibrium moisture content is calculated from RH and temperature using Equation \ref{eq:equil}. The fuel class, e.g. 10-hour fuel, characterizes how quickly the FMC responds to a change in atmospheric conditions. Figure \ref{fig:eq_plot} shows the FMC over time, along with the wetting and drying equilibrium moisture content. Notice how the upward sloping sections of the FMC curve occur slightly later in time than the upward sloping sections of the equilibrium moisture variables in Figure \ref{fig:eq_plot}. 

\subsection{Predictors Overview}

The set of predictors used in this project were chosen from physical considerations. The same set of predictors will be used for each ML model considered. Key physical drivers of FMC are RH, temperature, precipitation, wind speed, and solar radiation. All of these atmospheric variables are discussed previously in this report. In addition to these predictors, the hour of the day and day of year are used to help the models learn patterns in time. Finally, several variables that are meant to capture physical characteristics of the RAWS station locations are used as predictors. These include height above sea level, and the longitude and latitude coordinates of the stations. These predictors are used to help the model learn any spatial relationships with FMC.

Equilibrium moisture is derived from RH and temperature, so only equilibrium moisture will be used as it is generally inadvisable to include predictors that are direct functions of each other in ML models. However, further research could examine whether adding RH or temperature as predictors could improve the models. In physical models, the wetting and drying equilibrium variables are both used to capture physical behavior of fuels. In practice, these equilibrium values are very close to each other. In the ML context, these variables are so similar that it is preferable to just use one of them. For this project, the drying equilibrium will be used. 

All of the predictors used in this project, along with their units, are presented in Table \ref{tab:all_vars}. Basic summary statistics of these variables are presented in Table \ref{tab:all_dat_summary}.

\begin{table}[ht]
\centering
\caption{All Modeling Variables}
\label{tab:all_vars}
\begin{tabular}{|l|l|l|}
\hline
\textbf{Name}           & \textbf{Units} & \textbf{Description} \\  \hline
Equilibrium Moisture    & \%             & Derived from RH and temperature.    \\ \hline
Precipitation           & $mm/h$         & Calculated from all rain accumulated over the hour. \\ \hline
Wind Speed              & $m/s$          & Wind speed at 20ft above from.  \\ \hline
Solar Radiation         & $kWh/m^2$      & Downward shortwave radiation.   \\ \hline
Elevation               & feet           & Height above sea level.\\ 
\hline
Hour                    & hours          & From 0 to 23, UTC time.    \\ 
\hline
Day of Year             & days           & From 0 to 365 or 366\footnote{The data in this project included February 2024, which is a leap year month.}   \\ 
\hline
Latitude                & degrees        &  North to South spatial coordinate    \\ 
\hline
Longitude               & degrees        &  East to West spatial coordinate    \\ 
\hline

\end{tabular}
\end{table}

\begin{table}[ht]
\centering
\caption{Basic Summary Statistics for all Data}
\label{tab:all_dat_summary}
\begin{tabular}{llll}
\toprule
 & Min & Max & Mean \\
\midrule
Equilibrium Moisture & 1.04 & 38.06 & 17.75 \\
Precipitation & 0 & 70.61 & 0.05 \\
Wind & 0 & 41.13 & 2.66 \\
Solar Radiation & 0 & 1,245 & 185.84 \\
Elevation & 1,000 & 11,555 & 6,104.04 \\
Hour & 0 & 23 & 11.50 \\
DOY & 1 & 365 & 186.14 \\
Longitude & 37.09 & 45.92 & 40.96 \\
Latitude & -110.95 & -95.85 & -104.71 \\
\bottomrule
\end{tabular}
\end{table}

\section{Methods}

The purpose of the analysis described in this section is to analyze custom loss functions and evaluate whether they lead to an improvement in predictions of wildfire ROS. Several different ML models are used in order to ensure that the changes in prediction accuracy are related to the loss functions themselves, rather than properties of the particular model used. The ML models will be trained on observations of FMC, then used to predict FMC in the future and at unobserved locations. Finally, these FMC predictions will be transformed into theoretical wildfire ROS values, directly using the curve from Figure \ref{fig:fmc_ros_3wind}. These predicted ROS values will be compared to the ROS values calculated using the observed FMC to compute accuracy metrics.

\subsection{Machine Learning Models}
\label{sec:ml}

The goal of this project is to analyze the effect of using the custom loss functions for training ML models of FMC. Several different ML models will be considered in order to examine whether the custom loss functions have a similar effect across different model architectures. The models used in this project will be linear regression, XGBoost, and Random Forest. See Appendix \ref{app:ml} for more information on the models. These models have been tested by other researchers in the FMC modeling literature \citep{Lee-2020-EFM,McCandless-2020-EWS, Schreck-2023-MLV}. For each model, the same sets of training and testing data will be used. 

Each model will therefore be modeling the same functional mapping from the predictors at time $t$ to the response variable at time $t$. Let the response variable (FMC) at time $t$ be $y_t$, $X_t$ be the set predictors at time $t$, and $W$ be the set of model parameters. The training procedure selects the model parameters $W$ that minimize the loss function. So the general form of the models is:

\[
y_t = f(X_t, W)
\]

Hyperparameter selection is important for ML models. Hyperparameters in the context of ML are fixed parameters that are not updated in the training processes. Adjusting hyperparameters is known as ``tuning", and requires repeatedly training the model and evaluating forecast accuracy. However, it can be incredibly computationally intensive to effectively tune hyperparameters. It requires repeated training and testing steps, and additionally it requires constructing a ``validation set" from the data to avoid the situation where you select hyperparameters that overfit to the testing set. In this project a simple manual search was conducted. This will result in models that are slightly less accurate than if a more intensive hyperparameter search were conducted, but the advantage is that they are much quicker to deploy and analyze, and there is less of a necessity to use up data constructing validation sets. Future research will utilize a more systematic approach to hyperparameter selection. 

\subsection{Proposed Loss Functions}

The purpose of the custom loss functions proposed in this project is to place more weight on model residuals for drier fuels. It is hypothesized that this will lead to more accurate ROS forecasts relative to the standard MSE due to the nonlinear relationship between FMC and ROS, as depicted in Figure \ref{fig:fmc_ros_3wind}. A simple exponentially weighted loss function will be considered. From Equation \ref{eq:wmse}, the weights $w_i$ will be a function of the level of the $i^{th}$ observation of the response variable $y_i$. The weight for the $i^{th}$ residual and corresponding loss function will be:

\begin{equation}
    \label{eq:weights}
    w_i = \exp(-\omega y_i)
\end{equation}

\begin{equation}
    \label{eq:wloss}
    \text{Exponentially Weighted MSE} = \frac{1}{N}\sum_{i=1}^N w_i\cdot (y_i - \hat y_i)^2
\end{equation}

\begin{figure}[ht]
    \centering
    \includegraphics[width=1.1\textwidth]{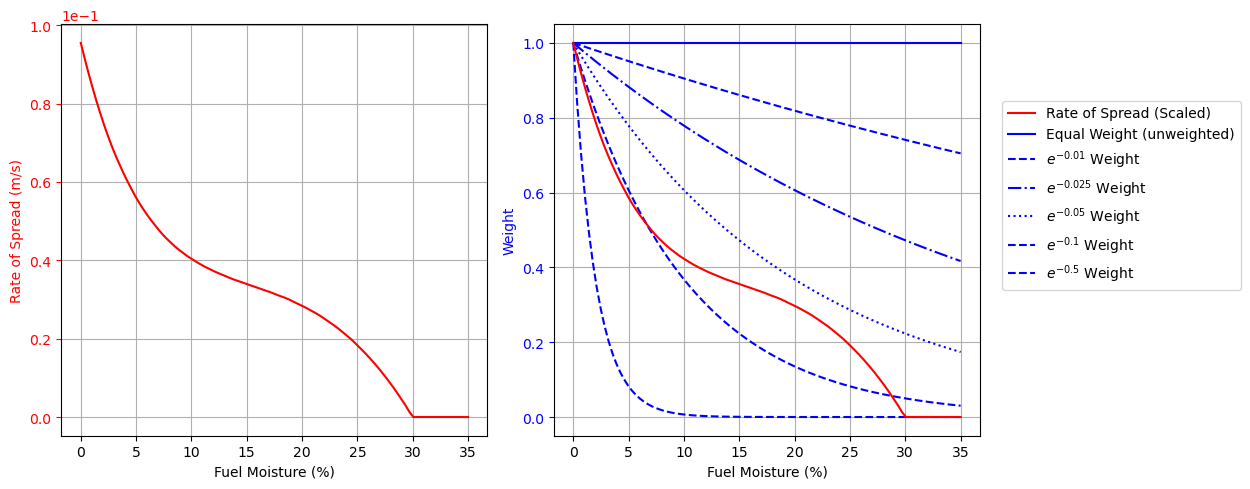}
    \caption{Loss Function Weights.}
    \label{fig:weights}
\end{figure}

The parameter $\omega$ represents the strength of the weighting scheme relative to an unweighted MSE. When $\omega = 0$, $w_i = 1$ for all $i$, and Equation \ref{eq:wmse} reduces to the standard MSE defined in Equation \ref{eq:mse}. As $\omega$ increases, model residuals associated with larger values of $y_i$ (wetter fuels) are multiplied by a smaller weight and thus contribute less to the overall model loss. Figure \ref{fig:weights} shows the weighting scheme for various values of $\omega$, ranging from $0$ to $0.5$. The line labeled ``Equal Weight" corresponds to the standard MSE and applies a weight of 1 for all residuals in the loss function. The dotted and dashed lines represent the weights used by the exponential weighting schemes. The $\omega$ parameter is the constant in the exponential equation, and as that parameter increases, less weight is given to residuals associated with higher observed values of FMC. The exponential weighting function is defined to be negative, so as $\omega$ increases, the exponential has a decreasing negative power. The ROS curve is plotted together with these exponential weighting curves for comparison.I n the right side of Figure \ref{fig:weights}, the ROS curve is rescaled for visual purposes.

Another way to construct a weighted custom loss function is to directly use the ROS value as a weight. Following the curve of the ROS based on FMC from Figure \ref{fig:fmc_ros_3wind}, assuming a constant 3m/s wind speed, a function called ``ROS" is derived which transforms FMC values into ROS. So for the $i^{th}$ observed value of FMC, the $i^{th}$ ROS value would be $ROS(y_i)$. This value can be used as a weight, and it has a similar shape to the negative exponential values, as seen in Figure \ref{fig:weights}. The ROS curve is strictly decreasing with respect to FMC, just like the negative exponential curves. A key mathematical difference between the ROS weights and the exponential weights is that the exponential curves will never assign a weight of zero. The ROS curve reaches zero at the moisture of extinction, or at 30\% FMC in Figure \ref{fig:weights}, so zero weight will be given to the residuals associated with FMC observations of 30\% or greater. In practice, these extremely high values are rare and would only be seen following periods of sustained heavy rain. Equation \ref{eq:rloss} shows the mathematical form of the ROS weighted loss function.

\begin{equation}
    \label{eq:rloss}
    \text{ROS Weighted MSE} = \frac{1}{N}\sum_{i=1}^N ROS(y_i)\cdot (y_i - \hat y_i)^2
\end{equation}

For this project, the following loss functions will be considered:

\begin{itemize}
    \item The standard MSE.
    \item Exponentially weighted MSE with a grid of 10 different $\omega$ values, ranging from $0.01$ to $0.25$.
    \item A weighted MSE with weights coming from the ROS curve with 3m/s wind, as shown in Figure \ref{fig:fmc_ros_3wind}. The results with zero wind were also analyzed and show similar results.
\end{itemize}

As mentioned before, it is hypothesized that some amount of weighting will improve ROS predictions. If the weighting is too strong, or if $\omega$ is too large, it should reduce the accuracy of the predictions. The use of the weighting scheme derived from the ROS function is hypothesized to be the best-case scenario of the weighting scheme. The same ROS function will be used to transform FMC values, so it is expected that directly using this function to derive weights should result in the best forecasts. However, the ROS function makes some important assumptions, notably that there is a constant 3m/s wind speed and zero slope. In reality, the ROS would be a complicated function of the local environment and fuel properties. The exponentially weighted loss functions do not rely on these physical assumptions, and are expected to be valid in a broader set of circumstances than a particular ROS weighting scheme.

\subsection{Analysis Description}

To evaluate the effects of the various loss functions on the accuracy of ROS forecasts, an experimental setup was constructed where the three ML models were trained on a period of data and used to predict the FMC at unobserved times and locations. The FMC predictions were transformed into ROS values for final accuracy metrics. 

\subsubsection{Cross-Validation of Spatiotemporal Models}
\label{sec:cv}
\hfill

In statistics, cross-validation refers to a set of methods used for evaluating model accuracy on unobserved data \citep[Chap.\  7.10]{Hastie-2010-ESL}. The dataset used to fit model parameters is referred to as the ``training set" of data, and the dataset used to calculate prediction accuracy for the model is referred to as the ``test set" of data. It is important that the testing data be independent of the training data. If not, the situation is known as ``data leakage", and this can lead to unreliable estimates of model accuracy.

In the context of FMC modeling, the models are spatiotemporal in nature, or the models predict values at a specific time for a specific location in space. Cross-validation procedures must be formulated carefully to avoid data leakage. The typical approach in ML would be to take a random sample of all of the data and hold some portion out to use as a test set. This method would typically lead to data leakage in the spatiotemporal context. A simple random sample for cross-validation would result in a training set that includes observations from the same location as observations in the test set, resulting in data leakage in space since the training set is spatially dependent with the test set. Further, a simple random sample would result in a training set with observations that occur later in time than observations in the test set, resulting in data leakage in time since the predictions generated for the test set would not be true forecasts into the future. To get a meaningful estimate of model accuracy, the test set must consist of observations at locations not included in the training set and at times in the future relative to all of the training set. 

\begin{figure}[ht]
    \centering
    \includegraphics[width=1\textwidth]{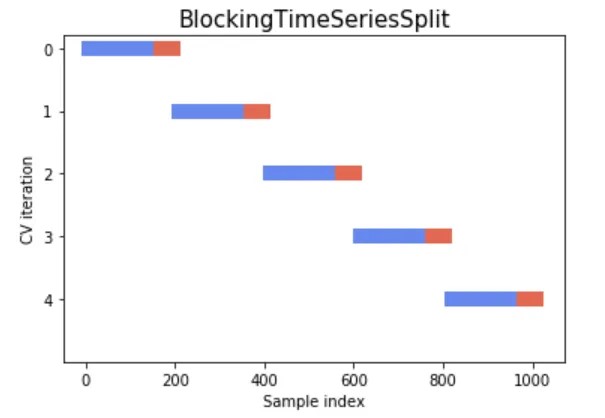}
    \caption[CV Method]{Cross-Validation with Blocked Subsets.\\ \textit{Reference: Soumya Shrivastava, Medium 2020.}}
    \label{fig:cv}
\end{figure}

To construct a test set at unobserved locations, a simple random sample of 20\% of the RAWS stations was selected for inclusion in the test set. So 80\% of the physical locations were used in each iteration of training. To account for the temporal relationships in the data when performing cross-validation, there are a number of possible techniques. In this project the method used was a version sometimes referred to as ``cross-validation with blocked subsets" \citep[p.\ 199]{Bergmeir-2012-OUC}. A fixed length of time is chosen for each iteration of training and testing. Figure \ref{fig:cv} shows a visual representation of the cross-validation method, when only considering time. In this Figure, the term ``sample index" on the x-axis corresponds to to the time, the blue blocks represent the training data, and the red blocks represent the testing data.

In this project, a training and testing time period will consist of 30 days; 28 days of training and 2 days (48 hours) of testing into the future. So starting at the earliest time record of the data,\footnote{After data filters, it was the time: May 17, 2023 at 2:22 UTC.} the first 30 days of observations are collected into the first time period. Using that 30 days of data, the random sample of locations will be performed as described above. Then the earliest 28 days are used as the training set, and the final 48 hours of observations from the 20\% subset of locations are used as the test set. So the test set will included observations at physical locations from the 20\% subset of locations, and at times post-dating all training observations. Thus, the test set consists of observations at locations not included in the training set and at times in the future relative to all data in the training set. In other words, the test set is independent of the training set, and there will be no data leakage. 

Models are fit to the training set, then predictions are made into the future on the test set, and finally accuracy metrics are calculated by comparing the forecasts to the observed values. After that, the 30 day time period window is moved 48 hours into the future. So the second iteration of test data were the 48 hours immediately following the previous 48 hours test set. In this way, the test sets had no overlap, but the training sets had at most 26 days of overlap. This led to a total of 167 training and testing periods. This was repeated 5 times to account to uncertainty, for a total of 835 periods of training and testing.

A testing period of 48 hours was selected since the HRRR weather model provides forecasts up to 48 hours into the future. Real-time wildfire simulations with the OpenWFM system will use HRRR weather data to make forecasts, and this project is intended to support the larger modeling efforts of OpenWFM. Thus in a real-time wildfire simulation, models could only be projected 48 hours into the future, and thus 48 hours is used as the length of the testing period in this project.

\subsubsection{Loss Function Experimental Design}
\hfill

Using the data collected from RAWS within the Rocky Mountain GACC, the data was split into training and testing periods using the process described in the previous section. A total of twelve loss functions were considered: the standard MSE, ten weighted schemes with varying $\omega$ parameters, and one with weights determined directly from the ROS transformation of FMC. For each of the twelve loss functions, the three ML models described previously were fit to the training set and the fitted model was used to predict new values in the test set. 

The model accuracy is calculated by comparing the predicted FMC in the test set to the observed FMC through the root mean squared error (RMSE). The RMSE is useful since it is interpretable in the same units as the response variable, so a RMSE of 4 can be interpreted as the model having on average a 4\% error in the FMC. The accuracy metrics for predicting FMC are computed to ensure that the levels of accuracy achieved for the models is close to what would be seen with models currently used in the wildfire modeling literature. The main goal, however, is to analyze the RMSE of predicted ROS values. The predicted FMC values were converted into theoretical ROS values through the curve seen in Figure \ref{fig:fmc_ros_3wind}, which assumes a constant 3m/s wind speed and zero slope. These predicted ROS values are compared to the ROS values calculated from the observed FMC in the test set.

For the 167 training and testing periods, 3 ML models were fit using 12 different loss functions. For each loss function, a total of 501 RMSE calculations were made across the time periods and models. These values were grouped together, so the final accuracy metrics for a particular loss function are a combination of RMSE values from different locations and different models. A total of 6,012 RMSE calculations were made across all models, time periods, and loss functions.

\subsubsection{Supplementary Analysis for Driest Period}
\hfill

When fuels are very dry, ROS can be expected to be higher. The effect of FMC on ROS will be the strongest when FMC values tend to be closest to zero. If custom loss functions are going to improve the accuracy of predicting ROS, that may be the most pronounced during the driest months with the highest risk for rapid fire spread. This section describes a supplementary analysis meant to analyze the driest period observed in the data. 

\begin{figure}[ht]
    \centering
    \includegraphics[width=1\textwidth]{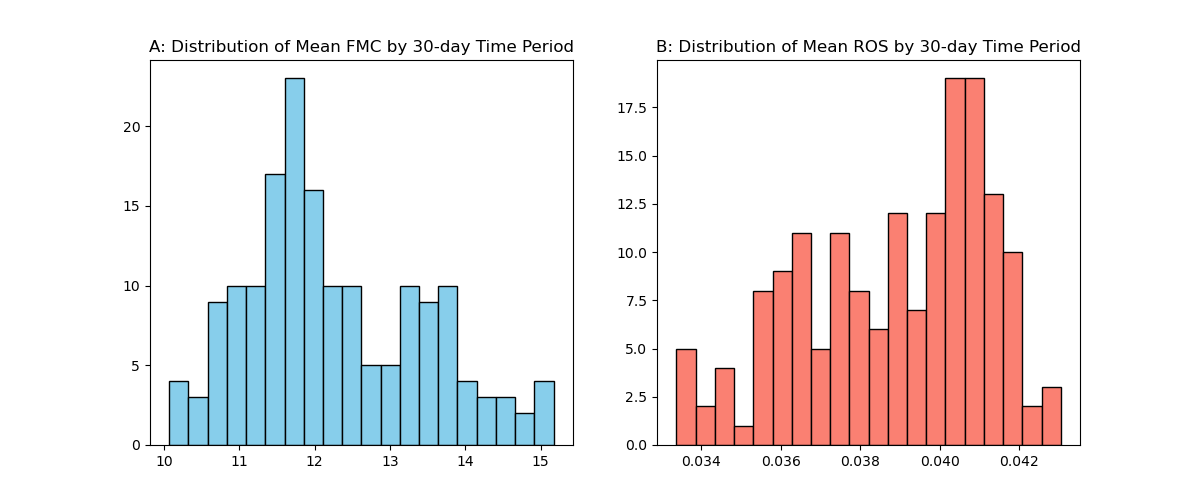}
    \caption{Time Period Histograms of FMC and ROS.}
    \label{fig:dry}
\end{figure}

Of the 167 time periods considered, the time period from October 18th, 2023 through November 17th, 2023 had the lowest average FMC and the highest average ROS, according to the constant 3m/s ROS function. See Figure \ref{fig:dry} for a summary of the average FMC and ROS for the time periods considered. It might be surprising that this period had the driest fuels in the year considered, but in the Rocky Mountain region weather conditions are very dry during those fall months. According to a summary of weather in 2023 from the National Weather Service for the weather around Denver, Colorado, that time period was very dry in terms of precipitation. November 2023 had the second lowest total precipitation (0.18 inches), just behind December when temperatures were also much lower \citep{NWS-2023-DAC}. The warmest months were July and August, but those months had far more precipitation than November  (2.10 and 0.93 inches, respectively). 

For this time period of data, the final 48 hours of data was used as the testing period, and the random splitting by physical location was repeated 10 times to account for uncertainty in the random sampling and variance in the random algorithms. Additionally, for computational efficiency, the loss functions considered were the standard MSE, the ROS weighting scheme, and only the weighted loss function with $\omega = 0.0367$, which was the best performing exponentially weighted loss function from the analysis over a full year.

\section{Results}

\subsection{Main Analysis - Over Whole Year}

The results of the analysis described above showed that there was a small improvement in the prediction accuracy for ROS for the custom loss functions across the set of three ML models.

To compare the RMSE of predictions from the models trained with different loss functions, the different time periods used for training and testing are considered to be independent realizations of the same data generating process. In other words, the RMSE for the various test periods and replications are aggregated to calculate a mean and standard error of the RMSE. When evaluating the prediction accuracy for FMC itself, the MSE loss function did not actually have the lowest RMSE on the test sets. Table \ref{tab:fmc_results} shows the RMSE values for predicting FMC aggregated over all models and time periods. The exponentially weighted loss function with $\omega=0.01$ and the ROS weighted loss function had the lowest mean estimates, very slightly lower than the standard MSE. 

\begin{table}[ht]
\centering
\caption{RMSE Results for predicting FMC.}
\label{tab:fmc_results}
\begin{tabular}{llrrr}
\toprule
 & Loss & Mean & Min & Max \\
\midrule
0 & MSE & 3.999 & 1.604 & 8.588 \\
1 & $exp\_0.01$ & 3.996 & 1.600 & 8.588 \\
2 & $exp\_0.0367$ & 4.000 & 1.418 & 8.786 \\
3 & $exp\_0.0633$ & 4.036 & 1.548 & 9.264 \\
4 & $exp\_0.09$ & 4.091 & 1.505 & 9.603 \\
5 & $exp\_0.1167$ & 4.157 & 1.449 & 10.226 \\
6 & $exp\_0.1433$ & 4.254 & 1.464 & 10.745 \\
7 & $exp\_0.17$ & 4.363 & 1.379 & 11.176 \\
8 & $exp\_0.1967$ & 4.489 & 1.382 & 11.531 \\
9 & $exp\_0.2233$ & 4.621 & 1.311 & 11.824 \\
10 & $exp\_0.25$ & 4.770 & 1.254 & 12.072 \\
11 & ROS & 3.996 & 1.573 & 8.817 \\
\bottomrule
\end{tabular}
\end{table}

When evaluating the prediction accuracy for ROS, the weighted loss functions show a small improvement in the RMSE. Table \ref{tab:ros_results} shows the RMSE values for predicting ROS aggregated over all models and time periods. The ROS weighted loss function had the largest improvement in accuracy, followed by the weighted loss function with $\omega=0.0367$. These were associated with reductions in the RMSE by 0.6\% and 1.1\%, respectively. Paired t-tests were use to examine the statistical significance of these mean differences. The paired t-tests compare the mean difference between the accuracy metrics from the standard MSE loss and the metrics from the two weighted loss functions. The assumed null hypothesis is that there is zero mean difference between the two groups of RMSE values, and a small p-value indicates evidence that the mean difference is nonzero. The p-values associated with the tests were roughly 0.017 when comparing the standard MSE to the exponentially weighted loss with $\omega=0.0367$, and roughly 0.005 when comparing the results for the standard MSE to the ROS weighted loss function. These are statistically significant at a relatively high threshold of significance, so this analysis provides reasonably strong evidence that the mean RMSE is less for the best custom loss functions when predicting ROS.

\begin{table}[ht]
\centering
\caption{RMSE Results for predicting ROS.}
\label{tab:ros_results}
\begin{tabular}{llrrr}
\toprule
 & Loss & Mean & Min & Max \\
\midrule
0 & MSE & 7.119e-03 & 3.039e-03 & 1.562e-02 \\
1 & $exp\_0.01$ & 7.105e-03 & 3.057e-03 & 1.524e-02 \\
2 & $exP\_0.0367$ & 7.074e-03 & 3.044e-03 & 1.566e-02 \\
3 & $exp\_0.0633$ & 7.080e-03 & 2.844e-03 & 1.651e-02 \\
4 & $exp\_0.09$ & 7.118e-03 & 3.127e-03 & 1.648e-02 \\
5 & $exp\_0.1167$ & 7.164e-03 & 3.062e-03 & 1.615e-02 \\
6 & $exp\_0.1433$ & 7.259e-03 & 3.093e-03 & 1.690e-02 \\
7 & $exp\_0.17$ & 7.374e-03 & 3.121e-03 & 1.759e-02 \\
8 & $exp\_0.1967$ & 7.531e-03 & 3.100e-03 & 1.821e-02 \\
9 & $exp\_0.2233$ & 7.690e-03 & 3.146e-03 & 1.876e-02 \\
10 & $exp\_0.25$ & 7.889e-03 & 3.165e-03 & 1.925e-02 \\
11 & ROS & 7.040e-03 & 3.032e-03 & 1.571e-02 \\
\bottomrule
\end{tabular}
\end{table}

\begin{figure}[ht]
    \centering
    \includegraphics[width=1\textwidth]{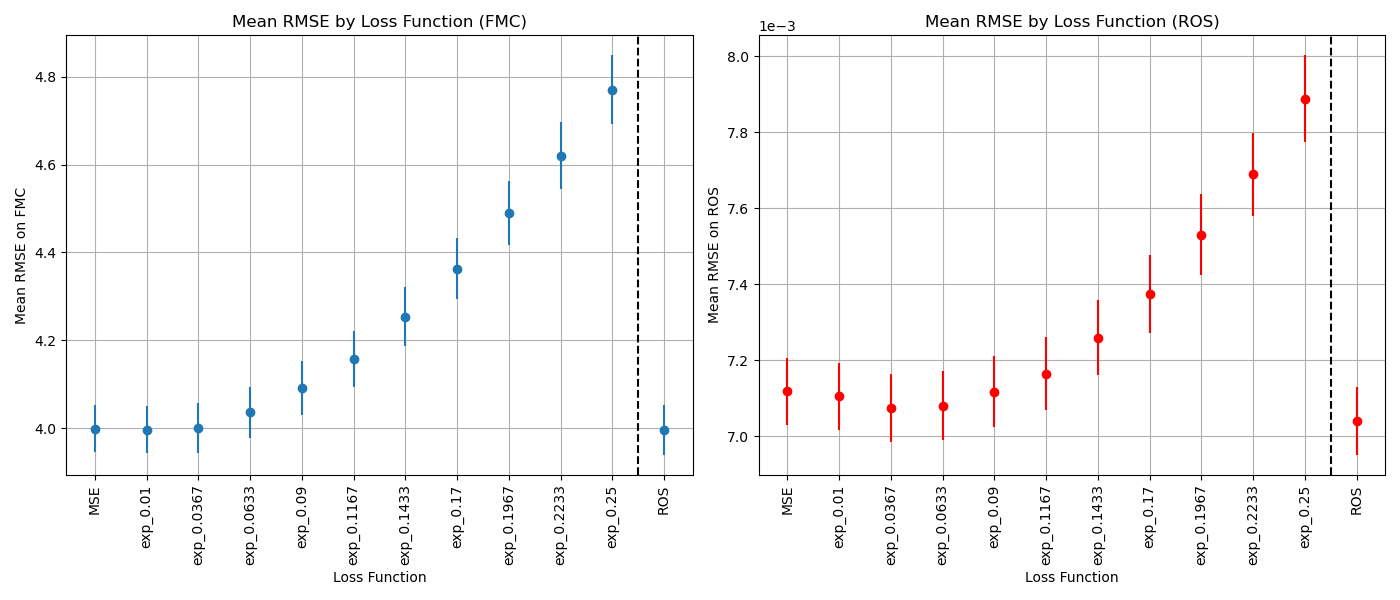}
    \caption{Forecast RMSE Results.}
    \label{fig:results1}
\end{figure}

Figure \ref{fig:results1} shows the results aggregated over the time periods. The dots represent the mean RMSE, and the vertical line the standard error of the mean RMSE. On the x-axis, the label ``exp\_0.01" corresponds to the exponentially weighted loss function with weight $w_i = \exp(-0.01\cdot y_i)$, where $y_i$ is the $i^{th}$ response value. The vertical dashed line in the plot is meant to distinguish the ROS weighting scheme from the others. The other loss functions are a continuum; the MSE corresponds to an $\omega$ parameter of zero, and then the $\omega$ parameter increases across the x-axis until ``exp\_0.25", which corresponds to $\omega=0.25$. The weighting scheme using the ROS curve is separate from this continuum. The left side of Figure \ref{fig:results1} shows the results of the trained models when predicting the untransformed FMC values. The right side shows the results of the models when predicting the transformed ROS response.

The results for predicting ROS show a U-shaped curve for the weighted loss functions, where a small amount of weighting of the residuals led to a smaller test RMSE, but as the strength of the weighting scheme increases the accuracy eventually deteriorates. The ROS weighting scheme resulted in the smallest test RMSE for ROS predictions, followed by the exponentially weighted loss with $\omega = 0.0367$. These results represent slightly over a 1\% reduction in the RMSE compared to the standard MSE loss function. Figure \ref{fig:results2} in Appendix \ref{app:res} shows the results broken down by ML model. The pattern is similar for the different models, supporting the idea that conclusions can be drawn about the loss functions themselves. For the XGBoost and linear regression models, the exponentially weighted loss with $\omega = 0.0367$ outperformed the standard MSE, but for the Random Forest the accuracy metrics were the lowest for the ROS weighted loss, but the exponentially weighted loss showed slightly higher test RMSE values. 

\subsection{Supplementary Analysis - Driest Period}

In this driest time period, the models are more accurate at predicting FMC than for the analysis over a whole year. This is because the spikes in FMC associated with rain events are much rarer in this time period. Table \ref{tab:fmc_results_dry} shows the RMSE for predicting FMC for the various loss functions across the 10 replications. Note that these are much lower than the values across the year shown in Table \ref{tab:fmc_results}, and again the FMC predictions were more accurate with the custom loss functions.

\begin{table}[ht]
\centering
\caption{RMSE Results for predicting FMC, Driest Month.}
\label{tab:fmc_results_dry}
\begin{tabular}{lrrr}
\toprule
Loss & Mean & Min & Max \\
\midrule
MSE & 2.507 & 1.758 & 3.334 \\
$\text{exp}\_0.0367$ & 2.433 & 1.711 & 3.232 \\
ROS & 2.408 & 1.693 & 3.211 \\
\bottomrule
\end{tabular}
\end{table}

The improvement in accuracy at forecasting ROS was much more pronounced than for the analysis run over the entire year. Table \ref{tab:ros_results_dry} shows the RMSE results across the 10 replications. The exponentially weighted loss function was associated with relative reduction of 3.4\% in predicting ROS relative to the standard MSE. The ROS weighted loss function showed a 4.7\% reduction. Paired t-tests for the mean differences relative to MSE were both well below 0.01, indicating a high level of statistical significance.\footnote{P-value for the exponentially weighted loss relative to the MSE was 3.68e-6, and for the ROS it was 3.93e-8} 

\begin{table}[ht]
\centering
\caption{RMSE Results for predicting ROS, Driest Month.}
\label{tab:ros_results_dry}
\begin{tabular}{lrrr}
\toprule
Loss & Mean & Min & Max \\
\midrule
MSE & 5.862e-3 & 4.091e-3 & 8.082e-3 \\
$\text{exp}\_0.0367$ & 5.665e-3 & 3.913e-3 & 7.539e-3 \\
ROS & 5.586e-3 & 3.886e-3 & 7.401e-3 \\
\bottomrule
\end{tabular}
\end{table}

\section{Discussion and Conclusions}

The results for the RMSE of the FMC values was unexpected. The MSE as a loss function resulted in the lowest MSE in the test set, but the differences in mean with the weighted loss functions were very small. Using biased estimators by weighting the loss function resulted in slightly lower RMSE on the test set. A hypothesis to explain this relates to the effect of rain. As was shown in Figure \ref{fig:fmc_with_rain}, following rain events the FMC spikes up and then quickly falls back down after the rain passes. These events are relatively rare in the data as there are not long sustained periods of nonzero rain. These spikes in FMC following rain are associated with high levels of FMC, which correspondingly receive less weight when using the weighted loss functions. So, the weighted loss functions may have the effect of downweighting ``outlier" events and causing them to have less of an effect on the model predictions. This finding may show that downweighting FMC observations following rain might lead to an overall more accurate model.

This analysis shows that weighted loss functions can produce at least as accurate results as a standard MSE loss function for ML models of FMC, when looking at both the prediction accuracy of FMC itself. And when predicting the transformation to ROS, the weighted loss functions show small but nonzero improvements in accuracy across the whole year of data. If the estimate of a roughly 1\% reduction in error of ROS were true, that might provide a meaningful improvement in wildfire simulations. When examining the driest month of data in the supplementary analysis, the predictions of ROS were much better relative to the standard MSE for the custom loss functions. This is encouraging for the idea that weighting the loss functions could improve real-world wildfire modeling, since the driest months are the periods when wildfire risk is the highest. The ROS weighted loss reduced the RMSE for predicting ROS by 4.7\%, and the best exponentially weighted loss reduced the RMSE by 3.4\%. If this improvement in accuracy held up for real-world wildfire models, that could lead to substantially more accurate predictions of wildfire behavior.

A number of simplification were used in this project. In real-time simulations of wildfires, ROS is a complicated function of local environmental conditions, such as terrain slope. It is not clear how weighted loss functions would perform under more realistic conditions for estimating ROS. Further, the relationship between FMC and ROS will be different for different fuel types. This project utilized the relationship for Anderson fuel category 8, closed timber litter. This fuel type is not purely 10-hour dead fuels. The ML models used in this report were also not subjected to exhaustive hyperparameter search schemes, and more finely tuned models might interact with the loss functions differently. Additionally, the models used in this report were static, meaning the time dependence of the system was not explicitly modeled. Also, this project used roughly 1 year of data from the Rocky Mountain region, and a more thorough analysis with data from other time periods and regions of the country could be performed to draw stronger conclusions about the effect of the weighted loss functions.

\setcounter{footnote}{0} 

Future research could apply the weighting scheme to RNNs or autoregressive models to examine the effect on forecast accuracy. One attractive feature about the weighted loss functions used in this project is that they are easy to implement in modern ML software packages.\footnote{The neural network software library Tensorflow, for example, has a parameter ``sample\_weight" that easily adds weighting to the standard MSE loss function.} Loss functions that can be framed in terms of a single weight applied to each residual are relatively easy to implement in code. A longer-term goal would be to run multiple real wildfire simulations, using a platform like WRF-SFIRE, with different estimates of FMC produced by models with weighted loss functions and see if wildfire behavior is captured more accurately.

In summary, this analysis provides evidence that custom loss functions which place more importance on the drier regions of model errors could improve predictions of wildfire rate of spread. Repeated training and testing periods resulted in reductions in the RMSE for predicting ROS by roughly 1\% to 5\%. This analysis justifies examining the effect of the weighted loss functions on larger datasets, more physically reasonable recursive models, and in real-world wildfire modeling.

\newpage

\begingroup
\raggedright
\bibliographystyle{ametsoc}
\bibliography{ref/fire,ref/fmc,ref/ml}
\endgroup

\appendix
\addcontentsline{toc}{section}{Appendices}
\addtocontents{toc}{\protect\setcounter{tocdepth}{1}}
\section{Data Processing} 
\label{app:data}

The data retrieval was performed on the University of Colorado Denver computing cluster. The retrieval code exists on github as part of the ``wrfxpy" python software package, part of the larger OpenWFM project. The exact python function, which was called with parameters from an executable file, can be found at: \url{https://github.com/openwfm/wrfxpy/blob/develop-72-jh/src/ingest/build_raws_dicts.py}

The code for the data processing can be found on github at: \url{https://github.com/jh-206/Custom-Loss-FMDA}

\subsection{Data Transformations}

Air temperature observations were converted from Celsius into Kelvin, since the constants in the equations of equilibrium moisture content depend on those units. Hourly rainfall, in units of mm/hour, was calculated from the hourly accumulated rainfall by taking the first difference in time. RAWS stations measure hourly accumulated precipitation using a rain gage bucket \citep{Campbell-2017-RMM}. These buckets fill up with rain over time and must be occasionally emptied. Some potential sources of error for this data include faulty sensors, uneven rain gage buckets, full rain gages that aren't emptied out when more rain is falling, etc. Rainfall observations were filtered out and considered missing if they were over 50mm/hr or less than zero. Additional filters could be explored in future research.

The time of the RAWS observations as retrieved from Synoptic is in Universal Time Coordinated (UTC) format. From this time, the hour of the day, from 0 to 23, is extracted. The hour of the day is used to help the models learn the periodic effect of the diurnal wetting and drying cycles that fuels experience. Humidity and temperature, the main physical drivers of fuel moisture, both have a diurnal pattern of regular highs and lows. These periods correspond with the diurnal pattern of the sun in the sky, where daily maximums in temperature and minimums of humidity occur in the afternoon. A simplifying assumption made in this project is to treat all of these hours the same across a larger geographic region. To make the hour of the day more physically meaningful, the hour could be adjusted based on the position of the sun as it moves in the longitudinal direction. The geographic region in this study is the Rocky Mountain GACC, which spans from central Utah through Kansas in the West to East direction. This is a distance of almost 1,500 kilometers. At those distances, there may begin to be meaningful physical difference in the hour of the day for different locations. 

\subsection{Data Filters}

Fuel moisture observations were considered missing if they were observed to be less than 1\%, which is considered not physically reasonable. Additionally, FMC sensors periodically break and will return erroneous data for some period of time. It is difficult to identify all instances where data is corrupted. One erroneous pattern is the RAWS FMC sensor returns the same constant value for an extended period of time. To identify these cases, the first time difference of FMC was calculated for each unique RAWS location. It was determined that RAWS station ``SAWC2" had erroneous data during the time period of this sutdy and was entirely removed from the data.

\section{Machine Learning Models of FMDA} 
\label{app:ml}

\subsection{Linear Regression}

The linear regression model used in this project uses the standard form with a constant intercept term, parameters for each predictor variable, and a random error term. As is standard in statistics, the random error is assumed to be normally distributed with zero mean and unknown variance. Let the response value at time $t$ be $y_t$, the number of predictors be $p$, the $i^{th}$ predictor at time $t$ be $x^{(i)}_t$, the $i^{th}$ parameter be $\beta^{(i)}$, and the random error be the  independent (across all $t$) and identically distributed random variable $\epsilon$ with zero mean and unknown variance $\sigma^2$. The model thus has the mathematical form \citep[e.g.][p.\ 44]{Hastie-2010-ESL}:

\[
y_t = \beta^{(0)} + \sum_{i=1}^p \beta^{(i)} x^{(i)}_t + \epsilon, \qquad \epsilon\sim N(0, \sigma^2)
\]

\subsection{Tree-based Methods}

In machine learning, ``tree" based models are a type of architecture used for classification or regression tasks where the data is partitioned some number of steps, and the final model output typically uses a simple average of the response variables associated with the data partitions. Regression trees thus produce piecewise-constant outputs \citep[Chap.\ 10]{Hastie-2010-ESL}.

The term ``ensemble" in machine learning refers to using many different models and aggregating them together. There are several motivations for using ensembles, including reducing the variance of the estimator, and across most modeling tasks there are improvements in accuracy when using ensembles as opposed to single trees. There are two popular forms of tree-based ensemble models that are both used in this project: random forests and XGBoost. These models have extensive sets of hyperparameters, which will be discussed below. In this discussion, we will identify cases where certain ranges of hyperparameters increase the risk of overfitting. Correspondingly, other ranges of those same hyperparameters would thus increase the risk of underfitting. But for concision, only overfitting is discussed.

\subsubsection{Random Forest}
\hfill

Random Forests are a variety of ensemble learner used in both regression and classification tasks. For some predetermined number of trees, a regression tree is built using a bootstrap sample of the training data, subsetted to a random sample of the available predictors. The bootstrapping and random sampling of the predictors used within the model reduces the variance of the estimator. When bootstrap sampling and random feature subsetting is not used, the trees in the ensemble are highly correlated to each other and the model struggles to learn the relationships between the response variable and the predictors \citep[p.\ 587]{Hastie-2010-ESL}. Table \ref{tab:rf-params} shows the hyperparameters used in this project for the Random Forest model.

Other hyperparameters exist for Random Forest models, but if they are not listed here their default values in the software were used. The ``n\_estimators" hyperparameter determines the total number of regression trees used in the ensemble. Larger values indicate a more complicated and computationally intensive model, and larger values also increase the risk of overfitting. The ``max\_depth" hyperparameter controls the depth of the regression trees in the ensemble, and again larger values indicate a more complicated and computationally intensive model, and larger values increase the risk of overfitting. The ``min\_samples\_split" hyperparameter sets a lower bound on the number of observations required to make a data partition in any given tree, and higher values increase the risk of overfitting. So the algorithm won't even consider a data partition on an existing partition that has fewer than 2 observations. Similarly, the ``min\_samples\_leaf" hyperparameter sets the lower bound on the number of observations that are in a partition. So the algorithm won't make a data partition if it leaves zero observations in one of the partitions. The ``max\_features" hyperparameter determines what fraction of the predictor variables are used in any given regression tree. Larger values increase the risk of overfitting. Finally, the ``bootstrap" parameter determines whether or not a bootstrap sample, or random sample with replacement, of the training data is taken for each regression tree in the ensemble. Bootstrapping is used to reduce the correlation between the training data in each tree and thus reduce the variance of the final estimator. 

For more information on the Random Forest hyperparameters, see \url{https://scikit-learn.org/stable/modules/generated/sklearn.ensemble.RandomForestRegressor.html}

\begin{table}[ht]
\centering
\begin{tabular}{|l|c|}
\hline
\textbf{Parameter} & \textbf{Value} \\
\hline
n\_estimators & 50 \\
max\_depth & 8 \\
min\_samples\_split & 2 \\
min\_samples\_leaf & 1 \\
max\_features & 0.8 \\
bootstrap & true \\
\hline
\end{tabular}
\caption{Random Forest Hyperparameters}
\label{tab:rf-params}
\end{table}

\subsubsection{XGBoost}
\hfill

Boosting refers to a variety of ensemble learning methods where the individual ensemble members are very simple, so-called ``weak learners", and then an iterative scheme reweights the data on the observations that the weak learners perform poorly on. In the context of regression trees, a simple regression tree is fit to the data in the first iteration. Subsequent iterations re-weight the data associated with observations that have relatively large residuals. 

Extreme gradient boosting, or XGBoost, is an augmentation of gradient boosting that provides computational benefits and additional functionality. Computational benefits of XGBoost relative to standard gradient tree boosting include efficiently handling sparse data with many missing values and support for parallelization. Additional modeling features of XGBoost include regularization of the parameters (i.e. L2 penalty similar to ridge regression) \citep[p.\ 337]{Hastie-2010-ESL}. The software documentation can be found online at: \url{https://xgboost.readthedocs.io/}

Table \ref{tab:xgb-params} shows the hyperparameters used in this project for the XGBoost model. These hyperparameters were used based on a compromise between computational efficiency and model accuracy. In the exploratory testing phase of the model tuning, the hyperparameters arrived at by \cite{Schreck-2023-MLV} for their XGBoost model were considered, but they ended up being less optimal than this set of hyperparameters. It should be noted that Schreck et al. used a broader set of predictor variables, including reflectance bands from VIIRS satellites. 

The ``max\_depth" hyperparameter controls the maximum depth of a regression tree in the ensemble. Larger values imply a more complicated and more computationally intensive model, and larger values increase the risk of overfitting. The ``eta" hyperparameter is the learning rate, which controls how quickly the model updates fitted parameters in each iteration. The ``min\_child\_weight" hyperparameter sets a lower bound for how small the data partition can be. Smaller values of this hyperparameter again imply a more complicated and more computationally intensive model, and smaller values increase the risk of overfitting. The ``subsample" hyperparameter controls what fraction of the training data is randomly selected at each iteration of building a tree in the ensemble. This hyperparameter can range from 0 to 1, and larger values increase the risk of overfitting. The ``colsample\_bytree" is similar to the previous hyperparameter, but controls what fraction of the predictor variables are used in each iteration. The ``n\_estimators" hyperparameter is the total number of trees in the ensemble. Larger values imply a more complicated and computationally intensive model, and also larger values increase the risk of overfitting. Finally, the ``gamma" hyperparameter is similar to ``min\_child\_weight" in that it sets a lower bound on the decrease in the loss function required to make a data partition. Other hyperparameters exist for XGBoost models, but if they are not listed here their default values in the software were used. For more information on these hyperparameters, see: \url{https://xgboost.readthedocs.io/en/stable/parameter.html}

\begin{table}[ht]
\centering
\begin{tabular}{|l|c|}
\hline
\textbf{Parameter} & \textbf{Value} \\
\hline
max\_depth & 4 \\
eta & 0.1 \\
min\_child\_weight & 1 \\
subsample & 0.8 \\
colsample\_bytree & 0.9 \\
n\_estimators & 120 \\
gamma & 0.1 \\
\hline
\end{tabular}
\caption{XGBoost Hyperparameters}
\label{tab:xgb-params}
\end{table}

\section{Further Results}
\label{app:res}

\begin{figure}[ht]
    \centering
    \includegraphics[width=1\textwidth]{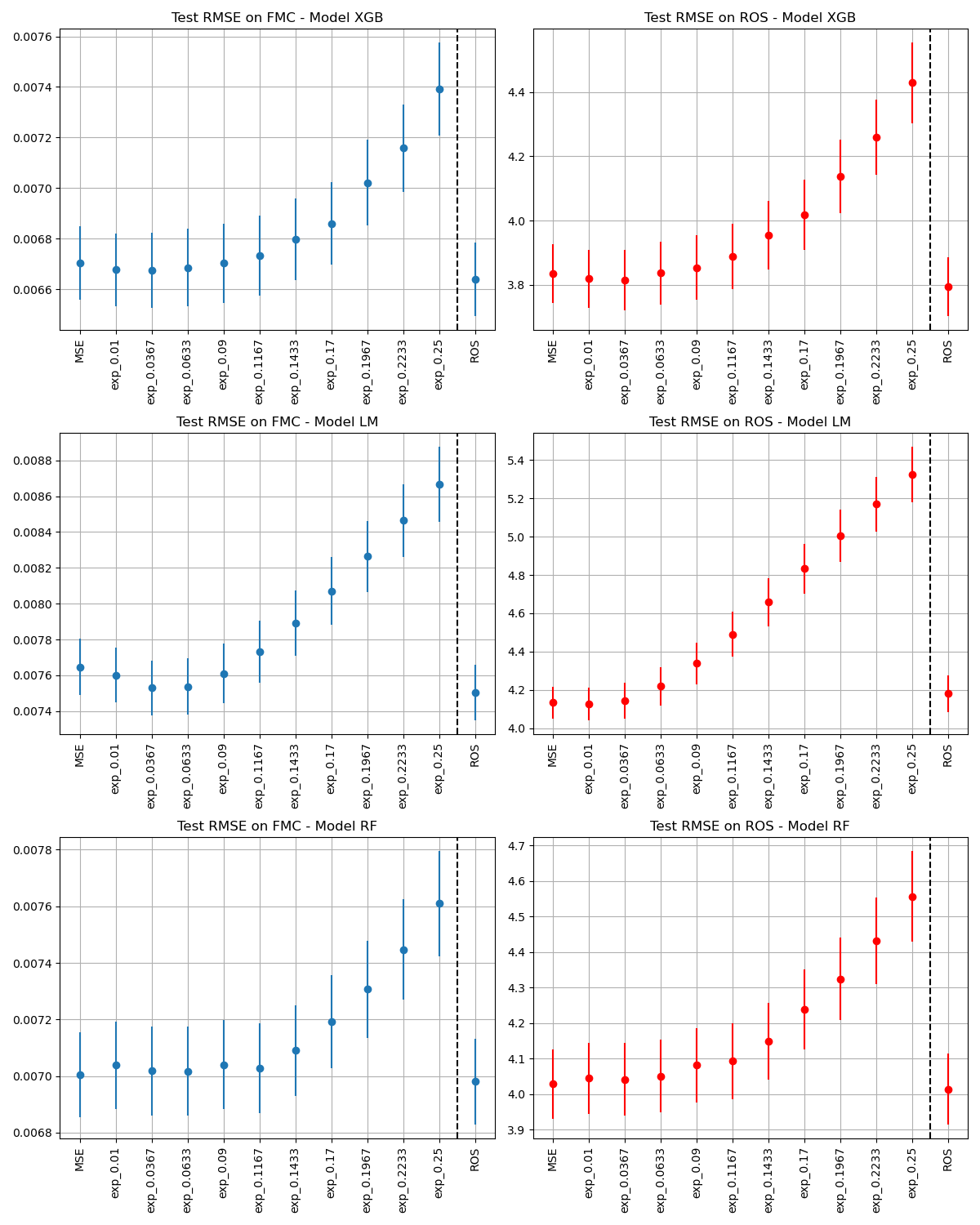}
    \caption{Forecast RMSE Results, by Model.}
    \label{fig:results2}
\end{figure}

Complete tables of results broken down by ML model can be found on github at: 

\url{https://github.com/jh-206/Custom-Loss-FMDA/blob/main/outputs/3_results-OUTPUTS.ipynb}

Figure \ref{fig:results2} shows the RMSE calculations for each of the ML models separately.

\end{document}